\documentstyle[twocolumn,pra,aps]{revtex}
\input epsf
\begin{document}
\draft
\title{Kinetic Roughening in Slow Combustion of Paper}
\author{M. Myllys$^{1}$,
        J. Maunuksela$^{1}$, 
        M. Alava,$^{2}$,  
        T. Ala-Nissila$^{2,3}$,
        J. Merikoski$^{1}$ and
        J. Timonen$^{1}$}
        
\address{
$^1$Department of Physics, University of Jyv\"askyl\"a,  
        P.O. Box 35, FIN--40351 Jyv\"askyl\"a, Finland \\
$^2$Laboratory of Physics, Helsinki University of Technology, 
        P.O. Box 1100, FIN--02015 HUT, Espoo, Finland \\
$^3$Helsinki Institute of Physics,  
        P.O. Box 9 (Siltavuorenpenger 20 C), FIN--00014  
        University of Helsinki, Finland, and Department of
        Physics, Brown University, Providence RI 02912--1843 \\
}

\date{11 May 2001, Submitted to Physical Review E}
\maketitle
 
%%%%%%%%%%%%%%%%%%%%%%%%%%%%%%%%%%%%%%%%%%%%%%%%%%%%%%%%%%%%%%%%%%%%%%%%%
%%                                                                     %%
%%  ABSTRACT                                                           %%
%%                                                                     %%
%%%%%%%%%%%%%%%%%%%%%%%%%%%%%%%%%%%%%%%%%%%%%%%%%%%%%%%%%%%%%%%%%%%%%%%%%
\begin{abstract}  
  Results of experiments on the dynamics and kinetic roughening of
  one-dimensional slow-combustion fronts in three grades of paper are
  reported. Extensive averaging of the data allows a detailed analysis
  of the spatial and temporal development of the interface
  fluctuations. The asymptotic scaling properties, on long length and
  time scales, are well described by the Kardar-Parisi-Zhang (KPZ)
  equation with short-range, uncorrelated noise.  To obtain a more
  detailed picture of the strong-coupling fixed point, characteristic
  of the KPZ universality class, universal amplitude ratios, and the
  universal coupling constant are computed from the data and found to
  be in good agreement with theory.  Below the spatial and temporal
  scales at which a cross-over takes place to the standard KPZ
  behavior, the fronts display higher apparent exponents and apparent
  multiscaling.  In this regime the interface velocities are spatially
  and temporally correlated, and the distribution of the magnitudes of
  the effective noise has a power-law tail. The relation of the
  observed short-range behavior and the noise as determined from the
  local velocity fluctuations is discussed.
\end{abstract}
%%%%%%%%%%%%%%%%%%%%%%%%%%%%%%%%%%%%%%%%%%%%%%%%%%%%%%%%%%%%%%%%%%%%%%%%%
%%                                                                     %%
%%  PACS                                                               %%
%%                                                                     %%
%%%%%%%%%%%%%%%%%%%%%%%%%%%%%%%%%%%%%%%%%%%%%%%%%%%%%%%%%%%%%%%%%%%%%%%%%


%%% 64.60.Ht Dynamic critical phenomena
%%% 05.40.+j Fluctuation phenomena, random processes, and Brownian motion
%%% 05.70.Ln Nonequilibrium thermodynamics, irreversible processes

%%%%%%%%%%%%%%%%%%%%%%%%%%%%%%%%%%%%%%%%%%%%%%%%%%%%%%%%%%%%%%%%%%%%%%%%%
%%                                                                     %%
%%  INTRODUCTION                                                       %%
%%                                                                     %%
%%%%%%%%%%%%%%%%%%%%%%%%%%%%%%%%%%%%%%%%%%%%%%%%%%%%%%%%%%%%%%%%%%%%%%%%%
\section{Introduction}

The phenomenon of kinetic roughening of driven fronts is abundant in
Nature. It has also become one of the paradigms in the physics of
non-equilibrium systems. The theoretical interest arises since, as in
critical equilibrium systems, the roughening behavior can be
understood in terms of scaling exponents and scaling functions
\cite{Bar95,Hal95,Mea98}. A classification of kinetic roughening
phenomena can be obtained by mapping the dynamics to an appropriate
Langevin equation which describes the interface dynamics, and also
includes a noise term with system specific correlations and magnitude
distribution.

The simplest non-linear interface equation related to kinetic
roughening is the celebrated Kardar-Parisi-Zhang (KPZ) equation
\cite{Kar86}, also related to the Burgers' equation of hydrodynamics.
The KPZ equation for a growing surface can be expressed in the form
\begin{equation}
  \frac{\partial h(x,t)}{\partial t}= \nu \frac{\partial^2
    h(x,t)}{{\partial x}^2} +  \frac{\lambda}{2} \left [
    \frac{\partial h(x,t)}{\partial x} \right ]^2 + F + \eta, 
  \label{KPZ}
\end{equation}
where $h(x,t)$ is the height of the interface, $F$ the driving force,
and $\eta$ denotes the noise affecting the interface. In the case
$\eta$ is short-range correlated in time and space, it can be
substituted asymptotically by white, thermal (Gaussian) noise:
$\langle \eta(x,t)\rangle = 0$ and $\langle \eta(x,t)\eta(x',t')
\rangle = 2D\delta(x-x')\delta(t-t')$. The KPZ equation contains a
linear diffusion term which imposes a surface tension through $\nu$,
and the non-linearity proportional to $\lambda$.  The essential
physics of the KPZ universality class arises from the presence of
additional symmetries, in particular a non-zero growth velocity
component in the direction perpendicular to the local surface
orientation.

The scaling properties as implied by the KPZ equation can be discussed
starting from e.g. the Family-Vicsek scaling \cite{Fam85}.  It
establishes a scaling relation for the surface width $w^2(L,t) \equiv
\langle \overline{(h - \bar{h})^2} \rangle$ as a function of time $t$
and system size $L$, such that
\begin{equation}
  \label{scal}
  w^{2}(t) = t^{2\beta} f(t/L^z) \propto
  \left\{
    \begin{array}{ll}
      t^{2\beta}  &  \mbox{, for $\xi \ll L$;} \\
      L^{2\chi} & \mbox{, for $\xi \gg L$.} 
    \end{array} 
  \right.  
\end{equation}
Here $\xi \sim t^{1/z}$ denotes a correlation length, which increases
with time until the finite system size induces a saturation.  The
overbar and brackets denote spatial and noise averaging, respectively.
Meanwhile three exponents have been defined: the growth ($\beta$) and
roughness ($\chi$) exponents, and the dynamical exponent
$z=\chi/\beta$.  For the KPZ equation with white noise (``thermal''
KPZ or TKPZ), the exponents can be established exactly in one spatial
dimension, and are $\beta = 1/3$, $\chi=1/2$, and $z=3/2$
\cite{For77,Kar86}.

The TKPZ universality class is distinguished also by other
characteristic properties, and not only by the scaling exponents.  The
{\em scaling function} $f(x)$ of Eq.~(\ref{scal}) is another landmark
of the KPZ equation \cite{Kar86,For77,Med89,Hwa91,Fog98}.  In the
steady state the KPZ interfaces are in fact equivalent to random
walks, since a fluctuation-dissipation theorem exists in $(1+1)$
dimensions, depending only on the effective noise strength and surface
tension, but not on $\lambda$. On the other hand, temporal
fluctuations on short time scales, and transient phenomena, do depend
on the nonlinearity. Thus the scaling function $f(x)$ and its
properties provide a distinct confirmation of the TKPZ roughening.

Consider the spatial ($x'$) and temporal ($t'$) two-point function of
the interface fluctuations $\delta h (x,t)$ around the mean interface
$\bar{h} (t)$
\begin{equation}
  \label{pcf2}
  C_{2}(r,t) = \langle \overline{[\delta h(x',t')-\delta h(x'+r,t'+t)]^2} 
  \rangle_{x',t'}.
\end{equation}
The equal-time limit $t=t'$ defines $G_2 (r)$, and for the (T)KPZ
universality class we have
\begin{equation}
  G_2 (r) \approx A r^{2\chi}.
\end{equation}
Here $A$ denotes an amplitude that depends \cite{Hal95} on the
parameters of the KPZ equation Eq.~(\ref{KPZ}), $A = D/\nu$.  Since
this amplitude describes the steady-state fluctuations, it is related
to the saturation width, and one finds \cite{Ama92a,Ama92b,Kru92} that
$w_{sat} = (A/12)^{1/2} L^{\chi}$.  Likewise, for the temporal
correlation function, defined through Eq.~(\ref{pcf2}) by taking
$r=0$, we have \cite{Hwa91,Ama92a,Ama92b,Kru92}
\begin{equation}
  C_{s,2} (t) \approx B t^{2\beta},
\end{equation}
where $B = \vert\lambda\vert^{2\beta} A^{\beta +1} R_G$ denotes a
dynamical amplitude and includes a universal constant $R_G$ that
characterizes the universality class of the system. Notice that $B$
depends on $\lambda$, signalling that $C_{s,2}$ is related to the
transient interface properties.  $R_G$ can thus be determined as a
{\it universal amplitude ratio}, $R_G =
B/(\vert\lambda\vert^{2\beta}A^{\beta +1})$. Other similar amplitude
ratios can also be defined \cite{Hal95}. The values of $R_G$, as
obtained \cite{Hwa91,Ama92a,Ama92b,Kru92} from mode-coupling theory
and simulations, are in the range $R_G = 0.63-0.71$.  The correlation
functions can as well be used to determine the {\it universal coupling
  constant} \cite{Hwa91} such that
\begin{equation}
  g^{*} = \frac{\lambda}{2} \left [ \frac{A}{B^{z/2}} \right
  ]^{1/\chi}.  
\end{equation}
Hwa and Frey have obtained \cite{Hwa91} the value $g^* \simeq 0.87$
based on a mode-coupling solution of the $(1+1)$-dimensional TKPZ
problem.  This calculation was later revised, and put in a more
systematic form \cite{Fre96} (see also \cite{Fog98}), and compares
well with the results from numerical studies of the scaling function
\cite{Kru92,Ama92a}.

Thus there are two distinct goals for an experimental study of
interfaces, expected to be in the TKPZ universality class. The
power-law behaviors of the bare interface correlations should enable
an estimation of the exponents $\chi$, $\beta$ and $z$. Then, by
augmenting these data with a determination of $\lambda$, one can probe
the scaling behavior beyond the scaling exponents by considering the
universal amplitude ratios and, possibly, the scaling functions of the
fluctuations.  Experimentally, the issue is complicated by the
existence of {\em cross-overs} from short-range behaviors to the
asymptotic TKPZ behavior, which would affect e.g. the behavior of the
height distribution, as compared to the pure TKPZ case.  We have
previously reported results on both the asymptotic scaling
\cite{Mau97} and on the properties of the short-range (SR) regime
\cite{Myl00}.  In this paper our goal is to present a coherent picture
of TKPZ physics in an experimental system, including more data and
discussion related to the behavior at short scales.  To this end, we
have carried out additional high-resolution combustion-front
experiments using very thin and light lens paper similar to that used
in \cite{Zha92}, in addition to our previous data \cite{Myl00} on two
grades of ordinary copier paper, which are much thicker and heavier.
The existence of large quantities of data allows us to do extensive
averaging, typically over $10000-40000$ interfaces. This is necessary
to establish the correlation functions and their scaling properties
reliably, in complete analogy with numerical experiments.  Our results
establish, clearly, the existence of asymptotic TKPZ scaling in terms
of the exponents. Moreover, we are able to determine the universal
amplitude ratios and the fixed-point coupling constant in fair
agreement with theoretical estimates.

The short-range behavior in an experimental system can differ from the
asymptotic scaling because of three fundamental reasons.  First, the
effective interface equation may become TKPZ-like only {\em during
  coarse-graining}. A case of this kind is provided by the
Kuramoto-Sivashinsky equation \cite{KS}, where TKPZ scaling can be
found only at large enough scales. Second, irrelevant terms in the
renormalization group sense, neglected in the KPZ equation, may
produce \cite{Nee97} short-range effects.  The third possibility is
that, still within the experimentally accessible window of spatial and
temporal scales, the effecive noise $\eta$ in Eq.~(\ref{KPZ}) proves
to be more complicated. Consider a slowly moving interface, in the
proximity of a pinning/depinning transition that takes place at a
critical $F_c$.  Since the interface moves in a landscape with
frozen-in or quenched inhomogeneities, $\eta$ will depend on $h(x,t)$.
For large enough driving forces, $F\gg F_c$, such noise correlations
vanish.  For intermediate $F$ one can define a cross-over length
scale, above which the behavior and noise correlations are TKPZ-like.
Below that scale, very close to $F_c$, the moving interface should
have an effective $\beta \simeq \chi \simeq 0.75$ \cite{Les96},
accompanied by apparent multiscaling.

The correlations that arise from quenched noise can be generalized to
temporal or spatial power-law correlations in the $\eta$
\cite{Med89,Kui99} with the correlator $\langle \eta(x,t)\eta(x',t')
\rangle = 2D\rho(x-x')\Delta(t-t')$. If the correlators $\rho$ or
$\Delta$ decay slowly enough, \textit{i.e.} algebraically, the TKPZ exponents
change \cite{Med89}.  The presence of cutoffs in the algebraic decay
of $\rho$ and $\Delta$ will induce again a cross-over to the TKPZ
scaling.  Similarly, a power-law amplitude probability distribution
function $P(\eta ) \sim \eta^{-(1+\mu)}, \eta \ge 1$, can change the
scaling behavior. For pure power-laws, there is a critical value of
$\mu_c \simeq 4$, below which the roughening behavior depends on
$\mu$. For $\mu = 3$, e.g., the $q^{\rm th}$ order spatial two-point
correlation functions have \cite{Myl00} short range roughness
exponents that change from about 0.76 ($q=1$) to 0.15 ($q=9$), while
the asymptotic behavior is self-affine, with $\chi \simeq 0.76$
\cite{Bar92}.

In the following we will therefore also elucidate the SR behavior, by
looking at the noise in the slow-combustion experiments, and how the
correlations of this noise and its amplitude distribution are
reflected in the SR scaling. Many different experiments on surface
growth, erosion, imbibition (paper wetting), fluid invasion, and also
slow-combustion fronts in paper, have failed to demonstrate asymptotic
TKPZ scaling, though one can of course ask whether all of these
problems should be described by a local interface equation
\cite{Dube}.  Exceptional fluctuations in the noise amplitude have
been observed in fluid-flow experiments \cite{Hor91b}, and were
reported earlier by us \cite{Myl00} for slow-combustion fronts.  The
roles of the crossover, also noted in fluid-flow experiments
\cite{Hor91a}, and in penetrating magnetic flux fronts in thin-film
superconductors \cite{Sur99}, and of the effective noise in confusing
the issue are to be noted. This may suffice as an explanation as to
why TKPZ behavior was not seen in the first similar experiment on
slow-combustion front roughening \cite{Zha92} in contrast with our
results.

Our impression is that in the slow-combustion fronts the SR physics
largely arises from the presence of ``non-ideal'' noise related to the
quenched impurities in the burning material, and to the natural
fluctuations due e.g. to gas flow during the combustion process. We
observe non-self-affine SR scaling, or multiscaling: the exponents
$\chi$ and $\beta$ depend on the order of the correlation function
from which they are measured. A high local front velocity is naturally
coupled to a local, steep interface gradient. At such locations the
interface propagates faster both parallel and perpendicular to the
interface orientation, since $\lambda$ is positive.  In the SR regime
the front dynamics is coupled to the non-standard noise, which is both
SR correlated and displays non-trivial amplitude fluctuations. The
interplay of usual KPZ physics with the noise properties results in
the absence of usual self-affine scaling. One of the origins of the
quenched SR noise may lie in the local concentration variations of the
potassium nitrate that is used to control the slow combustion. These
experiments have been done at concentrations large enough to ensure
that an eventual pinning regime is far away ($F \gg F_c$).

The paper begins in Section II with a detailed description of the
high-resolution experimental setup. Section III is devoted to an
analysis of the paper samples on which the experiments are done. In
Section IV we present results for the scaling exponents for all three
paper grades used in the experiments, starting with the asymptotic
long-range limit and touching then on the short-range behavior.  Noise
correlations are studied in detail for the lens paper, and compared
with those for the copier papers reported in Ref.~\cite{Myl00}.
Finally, we determine the universal amplitude ratios for all three
experimental cases in the asymptotic regime.  Section V comprises a
discussion about implications of our results to theory and to other
related experiments.

%%%%%%%%%%%%%%%%%%%%%%%%%%%%%%%%%%%%%%%%%%%%%%%%%%%%%%%%%%%%%%%%%%%%%%%%%
%%                                                                     %%
%%  EXPERIMENTAL SET-UP                                                %%
%%                                                                     %%
%%%%%%%%%%%%%%%%%%%%%%%%%%%%%%%%%%%%%%%%%%%%%%%%%%%%%%%%%%%%%%%%%%%%%%%%%
\section{Experimental setup}

\begin{figure}
  \epsfxsize=\columnwidth\epsfbox{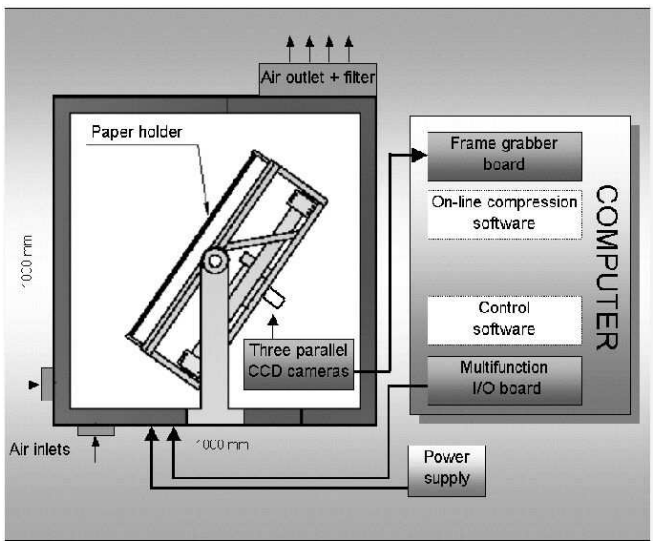}
  \caption{A schematic diagram of the experimental setup.}
  \label{fig:exp_setup}
\end{figure}

The experimental setup shown as a schematic diagram in
Fig.~\ref{fig:exp_setup} consists of a combustion chamber, a camera
system, and a computer with necessary hardware and software. One side
of the combustion chamber is made of glass and the opposite side is a
detachable aluminium plate for installing the paper samples. The rest
of the chamber sides are lined with a layer of porous material 50 mm
apart from the walls for making the incoming air flow laminar. In the
middle of the chamber there is a detachable sample holder, designed
for a maximum paper size of $600 \times 400$ mm$^2$. It can be rotated
with respect to the adjustable air flow so that convective transfer of
heat ahead of the propagation front can be regulated, and the other
flow-dependent features can be optimized. The sample holder is an open
metallic frame whose sides are both lined with needles that keep the
paper sheet planar during combustion. If the extra heat losses at the
boundaries need to be compensated for, the sides of the sample can be
heated with filaments that follow the combustion front. The volume
flow through the chamber was adjusted if necessary with an electrical
fan, placed on top of the chamber. The air flow through the chamber
was also numerically simulated, and the velocity field around the
sample was found to be quite homogeneous.  To minimize the dissipation
of heat from the combustion front, and to keep the air flow laminar,
we used relatively low volume flows (approximately 20 l/s for the
copier papers and no forced flow for the lens paper).

In the case the direction of air flow in the chamber was from bottom
to top, combustion fronts were ignited from the top end of the paper
samples by a tungsten heating wire in order to minimize convective
heat transfer. In the original setup reported in \cite{Mau97}, the
recording of the front was done by a single charge-coupled device
(CCD) camera, and the video signal was recorded on a Super VHS
recorder.  In the present \cite{Myl00} setup, the propagating front
was recorded with three parallel PULNiX TM-6EX black and white CCD
cameras, whose images are composed of $768 \times 548$ pixels. The
largest combined image contains $2304 \times 548$ pixels. Cameras were
attached to a stand that was operated with a pneumatic cylinder.  The
camera system was automatically moved in regular intervals along the
direction of propagation.

\begin{figure}
  \epsfxsize=\columnwidth\epsfbox{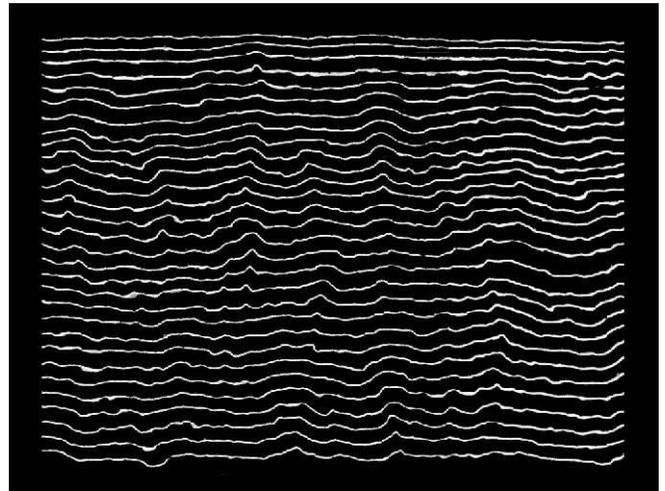}
  \caption{
    Series of typical digitised fronts.  The time step between
    successive fronts is 10 s, and the width of the digitised area is
    310 mm.}
  \label{fig:fronts}
\end{figure}

The digital frames of the three cameras were joined together and
compressed online, and then recorded on a hard disk using a
multi-level gray scale. The compressing of the frames was done by
recording only a narrow stripe around the front line. Since recording
was made in darkness, the only visible object was the combustion
front. By omitting the dark backround, we were able to reduce the size
of the individual frame file from 431 kB to approximately 15 kB. The
front height function was determined from each frame by first finding
the pixels brighter than a given gray-scale value. A single-valued
front line was fitted into the brightness profile in the strip found.
In the intervals, where a front line could not be identified, a
straight line was fitted by interpolation. The cylindrical image
distortions caused by the lenses were corrected using nonlinear
warping. The method needs a collection of 2D landmark points whose
true locations are known together with their distorted images. These
were then used to define a global warping function. Using this warping
function, corrections were made into the position data of the
individual cameras before joining them. A series of typical fronts
from the experiments are shown in Fig.~(\ref{fig:fronts}).

Improvements in data acquisition and introduction of the three-camera
setup increased the temporal and spatial resolution of the equipment.
For a 390 mm wide and 500 mm long paper sample, when the distance
between the sample and the cameras was 70 mm, the instantaneous
recorded area was $310 \times 74$ mm$^2$, and the pixel size was 0.135
mm.  This was an order of magnitude below the maximum typical length
scale related to areal mass variations in the paper samples, a few
millimeters. It was also of the same order of magnitude, actually a
bit smaller than the average width of the burning area in the
combustion fronts, which sets the lower limit to details that can be
detected. The maximum scanning frequency of the camera system was ten
frames per second. Since the average velocity of the fronts for the
copier paper used in the present experiments was approximately 0.5
mm/s, it was sufficient to store only every second frame of the
digitized fronts. During one time step, $\Delta t=0.2$ s, the front
propagated less than one pixel length.  For the lens paper we had to
slightly modify our experimental setup.  Needles in paper holder were
not suitable for attaching the thin paper, instead metallic bars were
used on both sides. Combustion fronts were ignited from the bottom and
every successive front was recorded because of their much higher
average velocity ($6-8$ mm/s) compared to those in copier papers.  nd,
as noted already above, for the lens-paper samples we used no forced
ventilation in the chamber.  This kind of setup was similar to that in
the early experiments reported in \cite{Zha92}.

%%%%%%%%%%%%%%%%%%%%%%%%%%%%%%%%%%%%%%%%%%%%%%%%%%%%%%%%%%%%%%%%%%%%%%%%%
%%                                                                     %%
%%  SAMPLES                                                            %%
%%                                                                     %%
%%%%%%%%%%%%%%%%%%%%%%%%%%%%%%%%%%%%%%%%%%%%%%%%%%%%%%%%%%%%%%%%%%%%%%%%%
\section{Samples}

The experiments were done using two different, easily available grades
of copier paper, with basis weights of 70 gm$^{-2}$ and 80 gm$^{-2}$,
and very thin lens paper with a basis weight of 9.1 gm$^{-2}$
\cite{papers}. It is important to realize that the composition of the
paper has an impact on the propagation of the combustion front.
Slow-combustion fronts do not easily propagate in a material made of
pure cellulose fibers only, because of their very low conductivity of
heat.  In normal copier papers approximately one fourth of the basis
weight is composed of fillers.  In our case, the fillers in both
grades of copier paper were mostly calcium carbonate (CaCO$_3$) with a
high heat capacity, in comparison with cellulose, which improved the
propagation of combustion fronts.  In addition, potassium nitrate
(KNO$_3$) was added as an oxygen source to all grades of paper to
ensure a uniform propagation of slow-combustion fronts. The
concentration of KNO$_3$ in the samples was kept at a value of
approximately 0.8 gm$^{-2}$.  Potassium nitrate was usually added by
embedding the samples in an aqueous solution of KNO$_3$ for five
minutes, after which they were dried in a press to maintain their
planar shape.  Due to the lower basis weight and much more porous
structure of lens paper, drying of KNO$_3$ solution was not
homogeneous over the sample, leading to an inhomogeneous concentration
distribution of KNO$_3$.  This would have a significant effect on the
experiments, as will be discussed later.  To avoid such problems,
KNO$_3$ was added to lens paper samples by spraying.

A second potential problem with paper samples is that it has been
shown \cite{Pro96} that, especially for low basis-weight laboratory
paper sheets, there may exist nontrivial power-law correlations in the
basis wight that may extend up to about 15 times the fiber length,
\textit{i.e.} into the centimeter range.  As it is well known
\cite{Tan92b} that correlated noise affects the scaling behavior of
kinetic roughening, maps of the local mass variations were prepared
for both grades of copier paper before and after the KNO$_3$
treatment. These $\beta$ radiographs were taken from paper samples of
size $170 \times 90$ mm$^2$ using spatial resolution of 0.04 mm.  For
the lens paper we used optically scanned images to estimate the local
mass variations.  In addition, we measured the calcium and potassium
concentration distributions in several layers of the same samples
using the laser-ablation method \cite{Hak95} with a spatial depth
resolution of 0.2 mm.

\begin{figure}
  \epsfxsize=\columnwidth\epsfbox{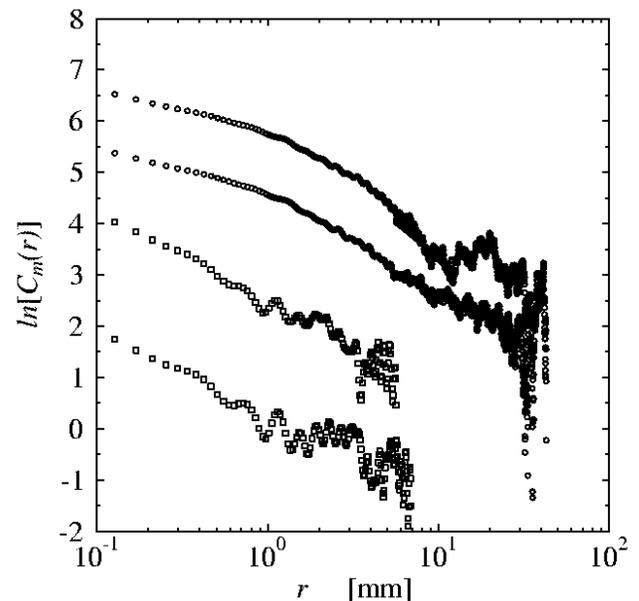}
  \caption{
    A log-log plot of the correlation function $C_{m}(r)$ for four
    copier paper sheets of basis weights 70 gm$^{-2}$, 70 gm$^{-2}$
    treated with KNO$_3$, 80 gm$^{-2}$, and 80 gm$^{-2}$ treated with
    KNO$_ 3$ (from top to bottom). The dashed line shows Gaussian
    behavior.  The curves have been shifted vertically for clarity.}
  \label{fig:radiog_cf}
\end{figure}

The radiographs of the copier paper samples were analysed using the
two-point density fluctuation correlation function
\begin{equation}
C_{m}(\vec{r})=\langle [m(\vec{x})-\bar{m}]
[m(\vec{x}+\vec{r})-\bar{m}]
\rangle, \label{2pc}
\end{equation}
where $m(\vec{x})$ is the local areal mass or the local basis weight,
$\bar{m}$ its average, and brackets denote an average over the sample.
The results in Fig.~\ref{fig:radiog_cf} show that in both cases, after
a distance of about one millimeter, a faster than algebraic
(power-law) decay of correlations can be seen.  This is in particular
the case with the 70 gm$^{-2}$ paper, while in the heavier paper
samples the noise level is reached rapidly.  We may conclude that
addition of KNO$_3$ has not affected the basis weight correlations.
The basis weight distributions $P(m)\equiv N(m)/\sum_{m'} N(m')$,
where $N(m)$ is the number of locations with the same basis weight,
differ from Gaussians only on the low basis-weight sides, which is
most likely caused by rounding errors in the digitization of the
$\beta$ radiographs.

The results of the laser-ablation measurements were analysed by a
correlation function similar to that in Eq.~(\ref{2pc}). The
difference was that, instead of the local basis weight, we used either
the calsium or potassium atomic emission line intensity, which is
proportional to the vaporized mass. The intensity-fluctuation
correlation functions were found to collapse to the noise level within
a distance of one millimeter in every layer. After the KNO$_3$
treatment the mean potassium intensity was found to be greater for the
70 gm$^{-2}$ than for the 80 gm$^{-2}$ paper. In order to reach the
same concentration of potassium nitrate for both grades, we used a
stronger aqueous solution of KNO$_3$ for the heavier paper.

To summarize, the results indicate that there are no correlations in
the filler, the local basis weight, or potassium nitrate distributions
beyond a few millimeters, and that the structures of the samples are
isotropic.

%%%%%%%%%%%%%%%%%%%%%%%%%%%%%%%%%%%%%%%%%%%%%%%%%%%%%%%%%%%%%%%%%%%%%%%%%
%%                                                                     %%
%%  KINETIC ROUGHENING OF COMBUSTION FRONTS                            %%
%%                                                                     %%
%%%%%%%%%%%%%%%%%%%%%%%%%%%%%%%%%%%%%%%%%%%%%%%%%%%%%%%%%%%%%%%%%%%%%%%%%

\section{Kinetic roughening of slow combustion fronts}

\subsection{Front width and correlation functions}

The most straightforward way to estimate the scaling exponents $\beta$
and $\chi$ for self-affine fronts is to use the scaling properties of
the interface width $w(L,t)$, as mentioned in the Introduction.
However, due to the large fluctuations in data when dealing with slow
combustion fronts, it is useful to consider the $q^{\rm th}$ order
two-point height-difference correlation functions
\begin{equation}
  \label{2pcf}
  C_{q}(r,t) = \langle \overline{[\delta h(x',t')-\delta h(x'+r,t'+t)]^q} 
  \rangle_{x',t'},
\end{equation}
where, again, $h(x,t)$ is the height of the front at point $x$ and
time $t$, and $\delta h(x,t) \equiv h(x,t)-\bar{h}(t)$, and the bar
denotes an average over a front while the brackets denote an average
over all configurations (fronts and burns). Through this quantity, one
can define the two functions
\begin{equation}
  G_{q}(r) \equiv C_{q}(r,0) \sim r^{\chi_q},
\end{equation}
and
\begin{equation}
  C_{s,q}(t) \equiv C_q(0,t) \sim t^{\beta_q},
\end{equation}
which thus provide estimates for the roughness and growth exponents.
In the saturated regime the functions $G_q(r)$ can be averaged over
all times (steady-state configurations), and $C_{s,q}(t)$ over all
spatial points.

Another quantity of interest for which fluctuations can be efficiently
averaged out, is the \textit{local} width of the propagating front
$w(\ell,t)$, defined as
\begin{equation}
  \label{w2l}
  w^{2}(\ell,t) = \langle \langle [h(x,t)-\langle h(x,t) \rangle_{\ell}]^2 
  \rangle_{\ell} \rangle,
\end{equation}
where the notation $\langle \rangle_{\ell}$ denotes spatial averaging
over all subsystems of size $\ell$ of a system of total size $L$. For
growing self-affine interfaces, the scaling exponents satisfy $\beta_q
= q \beta$ and $\chi_q = q \chi$, and the local width follows the
Family-Vicsek scaling relation \cite{Fam85} given by
\begin{equation}
  w^{2}(\ell,t) \sim \left\{
    \begin{array}{ll}
      t^{2\beta}  &  \mbox{, for $t \ll \ell^z$;} \\
      \ell^{2\chi} & \mbox{, for $t \gg \ell^z$.} 
    \end{array} 
  \right . 
\end{equation}
This provides another way to estimate the scaling exponents from
experimental data.

\subsection{Results for scaling exponents}

\subsubsection{Data analysis}

In order to reduce the influence of boundary effects on the data, an
area of width 270 mm taken in the middle of the recorded area of width
390 mm was used to calculate $h(x,t)$. The scaling exponents $\chi_q$
and $\beta_q$ were then determined by performing a linear
least-squares fit to the corresponding two-point height correlation
functions (Eq.~(\ref{2pcf})) in the scaling regime.  An
\textit{independent} estimate for the roughness exponent $\chi$ was
obtained from the local width.  The early-time behavior of the surface
width (Eq.~(\ref{w2l})) also gave an estimate for the growth exponent
$\beta$. Accurate determination of the scaling exponents was
complicated by the ``intrinsic'' width of the fronts caused also by
random structural inhomogeneities of the samples. Therefore, in the
spirit of the usual \textit{convolution ansatz}
\cite{Wol87,Ker88,For90} by which a random process affecting the front
but independent of the front dynamics will induce an additive constant
in the square of the front width, we performed linear least-squares
analyses of $\log_{10}[G_q(2r)-G_q(r)]=q\chi \log_{10}(r)+\mbox{\it
  const}$, and of $\log_{10}[C_{s,q}(2t)-C_{s,q}(t)]=
q\beta\log_{10}(t)+\mbox{\it const}$, to get rid of additive constant
factors in the correlation functions. The results given in the text
are obtained after these ``intrinsic widths'' were subtracted.

\subsubsection{Roughness exponent $\chi$}

\begin{figure}
  \epsfxsize=\columnwidth\epsfbox{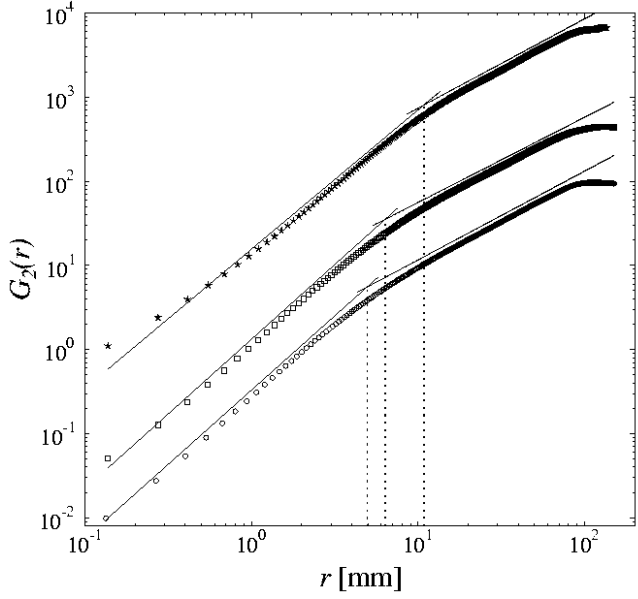}
  \caption{{The spatial correlation function}
    $G_{2}(r)$ {\it vs.} $r$ averaged over 10 burns of the 70
    gm$^{-2}$ ($\circ$) and 6 burns of the 80 gm$^{-2}$ copier paper
    ($\Box$), and 32 burns of the 9.1 gm$^{-2}$ lens paper ($\star$).
    The crossover lengths $r_{\rm c}$ are $5.2(1)$, $7.9(2)$, and
    $11.1(1)$ mm for these paper grades, respectively, shown with
    vertical lines, and the solid lines indicate slopes that
    correspond to the exponents in Table \ref{tab:res}.  The curves
    have been shifted for clarity.}
  \label{fig:c2r} 
\end{figure}

\begin{figure}
  \epsfxsize=\columnwidth\epsfbox{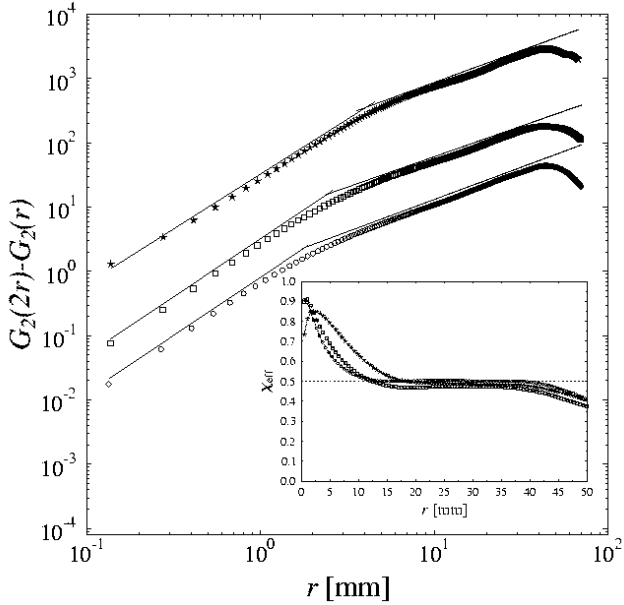} 
  \caption{{The spatial correlation function after the intrinsic width
      has been subtracted} $G_{2}(2r)-G_{2}(r)$ {\it vs.} $r$ averaged
    over 10 burns of the 70 gm$^{-2}$ ($\circ$) and 6 burns of the 80
    gm$^{-2}$ copier paper ($\Box$), and 32 burns of the 9.1 gm$^{-2}$
    lens paper ($\star$). The solid lines indicate slopes that
    correspond to the experiments in Table \ref{tab:res_2}.  The
    curves have been shifted for clarity. Inset shows the effective
    rougheness exponent $\chi_{\rm eff}$ {\it vs.} $r$, where the
    horizontal line is $\chi=1/2$.  }
  \label{fig:c2r_iw} 
\end{figure}

\begin{figure}
  \epsfxsize=\columnwidth\epsfbox{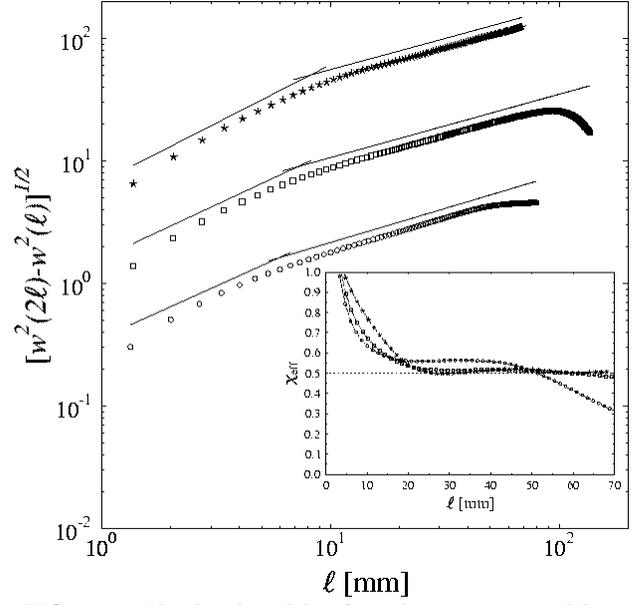}
  \caption{
    The local width after the intrinsic width was subtracted
    $[w^{2}(2\ell)-w^{2}(\ell)]^{1/2}$ {\it vs.}  $\ell$ averaged over
    11 burns of the 70 gm$^{-2}$ ($\circ$) and 18 burns of the 80
    gm$^{-2}$ copier paper ($\Box$), and 24 burns of the 9.1 gm$^{-2}$
    lens paper ($\star$). Solid lines indicate slopes that correspond
    to Table \ref{tab:res_2}. The curves have been shifted for
    clarity. Inset shows the effective rougheness exponent $\chi_{\rm
      eff}$.}
  \label{fig:w2l_iw}
\end{figure}

First, we present results for the roughness exponent $\chi_2$ using
the second-order correlation functions and the local width. The main
results for both $\chi = \chi_2/2$ and $\beta = \beta_2/2$ are
summarized in Tables \ref{tab:res} and \ref{tab:res_2} for all three
paper grades. In the latter Table all the fits have been done by first
subtracting the intrisic widths from the data.  The corresponding
spatial correlation functions $G_2(r)$ are shown in Figs.
\ref{fig:c2r} and \ref{fig:c2r_iw} so that in the latter figure the
intrinsic widths have been subtracted.  It is immediately evident from
the data that there are two regimes of apparent scaling, separated by
a crossover length $r_c$. In the SR regime below the crossover length,
$G_2(r)$ scales with a rather large effective exponent of $\chi_{\rm
  SR} \simeq 0.90$ for the heavier paper grades, and $0.85$ for the
lens paper.  In the long-range (LR) regime $r \gg r_c$, the effective
exponent $\chi_{\rm LR}$ converges very close to the exact KPZ value
of 1/2 for all three cases. Beyond about 100 mm the statistics becomes
worse and fluctuations larger.  The inset in Fig.  \ref{fig:c2r_iw}
shows estimates for the effective running exponents defined as
$\chi_{\mathrm{eff}}=(1/2)\log_{10}[(G_2(2r_2)-G_2(r_2))/
(G_2(2r_1)-G_2(r_1))]/\log_{10}(r_2/r_1)$, where $r_2=4\,r_1$.  In all
three cases, the LR asymptotic approach towards the TKPZ behavior is
evident. It is also evident from this inset that no clear scaling
regime can be found at short range with the higher apparent exponent.
The behavior of the local width $w^2(\ell)$ averaged over the same
data is consistent with $G_2(r)$ as is demonstrated in
Fig.~\ref{fig:w2l_iw} where the intrinsic width is subtracted.

\subsubsection{Growth exponent $\beta$}

\begin{figure}
  \epsfxsize=\columnwidth\epsfbox{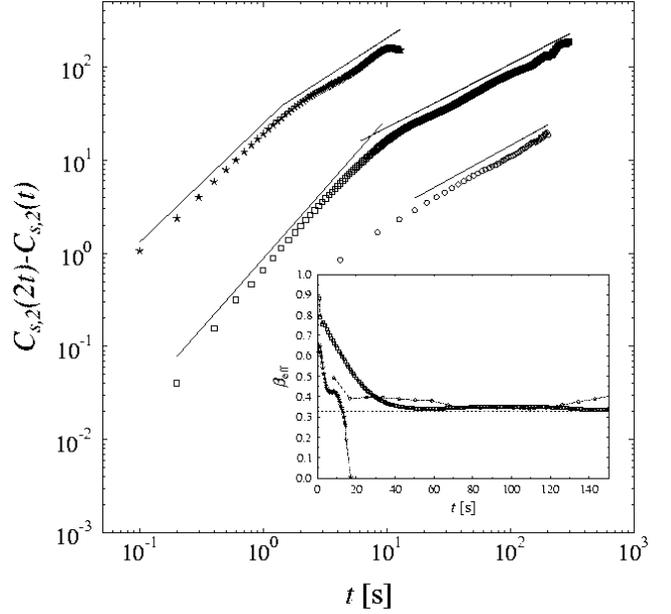}
  \caption{The temporal correlation function after the intrinsic width
    was subtracted $C_{\mathrm{s},2}(2t)-C_{\mathrm{s},2}(t)$ {\it
      vs.} $t$ averaged over 11 burns of the 70 gm$^{-2}$ ($\circ$)
    and 18 burns of the 80 gm$^{-2}$ copier paper ($\Box$), and 24
    burns of the 9.1 gm$^{-2}$ lens paper ($\star$). The curves have
    been shifted for clarity. The solid lines indicate slopes that
    correspond to the experiments in Table \ref{tab:res_2}. Inset
    shows the effective growth exponent $\beta_{\rm eff}$ {\it vs.}
    $t$. The horizontal line denotes $\beta=1/3$.}
  \label{fig:c2t_iw} 
\end{figure}

The temporal correlation functions $C_{2,s}(t)$ for the three cases
are shown in Fig.~\ref{fig:c2t_iw} with the intrinsic widths removed.
The poor time resolution ($\Delta t=$4.2 s) in the 70 gm$^{-2}$ paper
data prevented us from analysing in this case the temporal behavior
very accurately.  In the long-time regime the scaling behavior of the
paper grades is again in agreement with the TKPZ case. The measured
$\beta_{LR}$ is, in particular after the intrinsic width has been
removed, in good agreement with the TKPZ value 1/3.  The largest
deviations are shown by the lens paper as in this case the saturated
regime was too short for an accurate determination of $\beta$. As was
done in the case of the $\chi$'s, we also used the local width,
Eq.~(\ref{w2l}), to compute the $\beta_{LR}$. The scaling range is in
this case slightly less than a decade, and the exponents obtained
(without subtracting the intrinsic width) agree well with the TKPZ
one. Note that the same data for $ w(\ell,t)$ is used below to study
its scaling function.

In the data, there is a crossover at time $t_c$ from short-time
behavior to asymptotic long-time regime visible in each case. The
short-time growth exponent $\beta_{\rm SR}$ is again rather large,
$\beta_{\rm SR}=0.75(5)$ for the 80 gm$^{-2}$ case, and 0.64(3) for
the lens paper. In the long-time regime beyond the crossover time, the
exponent $\beta_{\rm LR}$ is again fully consistent with the TKPZ
value of 1/3.  The inset in Fig.~\ref{fig:c2t_iw} shows the approach
of the effective running exponents $\beta_{\mathrm{eff}} =
(1/2)\log_{10}[(C_{2,s}(2t_{2})-C_{2,s}(t_{2}))/
(C_{2,s}(2t_{1})-C_{2,s}(t_{1}))]/\log_{10}(t_2/t_1)$, where
$t_2=4\,t_1$, towards the TKPZ limit. Also in this case there is no
clear scaling regime below the crossover time, although an apparent
scaling exponent can be determined.

\subsubsection{Multiscaling at short range}

\begin{figure}
  \epsfxsize=\columnwidth\epsfbox{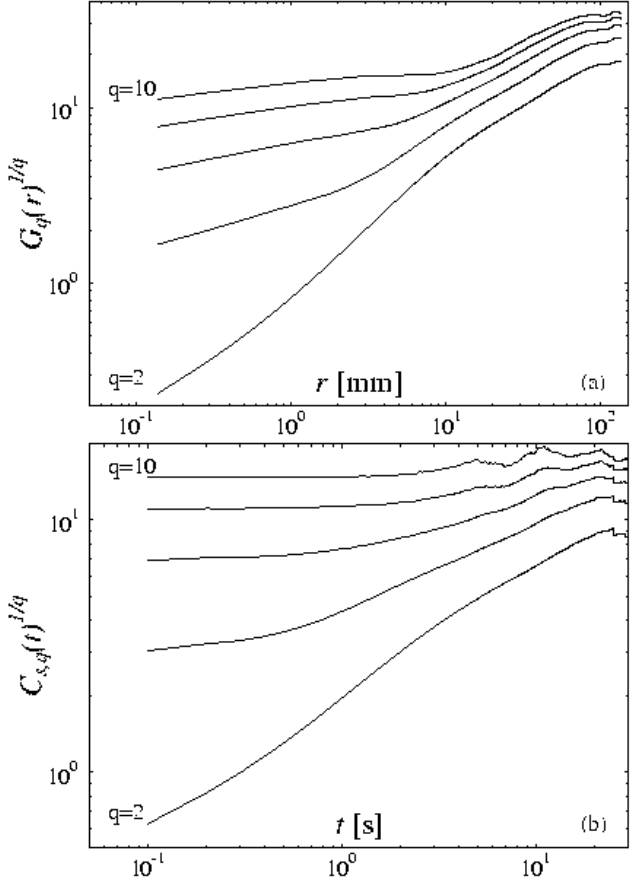}
  \caption{(a) The $q^{\rm th}$ order spatial correlation function
    $G_{{q}}(r)^{1/q}$ {\it vs.} $r$ averaged over 32 burns of the 9.1
    gm$^{-2}$ lens paper. (b) The $q^{\rm th}$ order temporal
    correlation function $C_{{s,q}}(t)^{1/q}$ {\it vs.} $t$ averaged
    over 24 burns of the 9.1 gm$^{-2}$ lens paper. The functions in
    (a) and (b) are plotted for even moments from $q=2$ to $10$.}
  \label{fig:cq}
\end{figure}

Spatial higher-order correlation functions, Eq.~(\ref{2pcf}), were
determined to check the possible multiscaling properties of the
combustion fronts for the lens paper, to accompany the data published
earlier \cite{Myl00} for the copier papers.  The behavior of the
higher-order correlation functions indicates that, in the TKPZ regime,
the interfaces are self-affine. On the other hand, in the SR regime
the slopes of the spatial and temporal correlation functions of
Figs.~\ref{fig:cq} depend on the order $q$ of the correlation
function, \textit{i.e.} show apparent multiscaling \cite{Bar91a,Bar91b,Bar92},
similar to that found for the two grades of copier paper \cite{Myl00}.

Analysis of the fitted SR scaling exponents reveals that the
$\chi_{SR}(q)$'s approach the value 1/2 for large $q$ which means that
larger local slopes follow a Gaussian distribution and are
uncorrelated as in the long-range TKPZ regime.  At the same time,
$\beta_{SR}(q)$'s become very small and seem to approach zero. We
conclude that the overall behavior of the SR exponents -- with some
uncertainty as for the higher momenta of the distributions -- is very
similar for all paper grades, and would thus seem to be a generic
feature in these systems. One should however realize that uncorrelated
random factors also appear in experiments, which affect the fronts and
especially their scaling properties at SR if they involve high
gradient values. They are not related to front dynamics and appear
therefore as ``artefacts''. Their effects can be analyzed
qualitatively by using standard filtering techniques (e.g. median
filtering).

\subsection{Noise correlations}

\begin{figure}
  \epsfxsize=\columnwidth\epsfbox{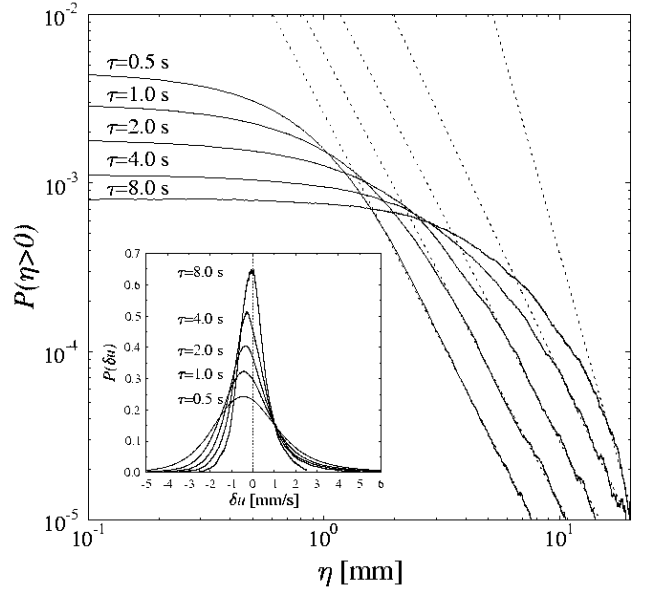}
  \caption{{The noise amplitude distribution} $P(\eta> 0)$ averaged 
    over 35 burns of the 9.1 gm$^{-2}$ lens paper with an average
    velocity $>8.4$ mm/s. The distribution was calculated for time
    intervals $\tau=0.5,1.0,2.0,4.0,$ and $8.0$ s shown in the
    figure. The linear fits have slopes $-2.71, -2.67,-2.72,-2.99,$
    and $-4.97$, left to right. The inset shows the local velocity
    fluctuation distributions for the same time intervals $\tau$.}
  \label{fig:pe}
\end{figure}

Next we consider the effective noise at the combustion fronts as
determined from the fluctuations in the front velocity similarly to
Ref.~\cite{Hor91b}. The physics behind this idea is that the presence
of local large velocities can be related to the bare noise properties
as the maximum front slopes are limited.  The effective noise can be
analyzed by considering the fluctuations of local velocities, defined
as $\delta u(x,t)=u(x,t)-\bar{u}(t)$, with $u(x,t)\equiv
(1/\tau)[h(x,t+\tau)-h(x,t)]$, which obviously depend on the timescale
$\tau$.  Because of the crossover behavior, one has to analyze the
noise correlations below ($\tau < t_c$) and above ($\tau > t_c$) the
crossover scale.  An alternative formulation for the noise amplitude
is
\begin{equation}
  \eta (x,t) \equiv \delta h(x,t+\tau)-\delta h(x,t),
\end{equation}
where $\delta h= h(x,t)-\bar{h} (t)$.  We have used this quantity
to estimate the amplitude distribution,
\begin{equation}
  P(\eta) \equiv \frac{1}{\sum_{\eta\,'} N(\eta\,')} N(\eta), 
\end{equation}
in the steady-state regime. Here $N(\eta)$ is the number of positions
where $\eta$ has the same value, and the sum goes over all values of
$\eta$.  We determined $\eta (x,t)$ for time steps
$\tau=0.5,1.0,2.0,4.0,$ and $8.0$ s for the lens paper, and show the
related distributions $P(\eta)$ for $\eta>0$ in Fig.~\ref{fig:pe}. For
short time steps the distributions have to a good degree of accuracy a
power-law tail of the form $P(\eta)=c\eta^{-(\mu+1)}$, with $\mu
\simeq 1.7$. For increasing $\tau$ the power-law contribution in the
tail of $P(\eta)$ becomes less visible, the exponent $\mu$ increases
towards $\mu = 5$, and the distribution approaches a Gaussian. The
behavior is similar to that found for the copier paper \cite{Myl00},
except for the exact value of the exponent $\mu$ which was found to be
about 2.7 at short times ($\tau=1.0$ s).

The local velocity fluctuations were calculated using time steps
$\tau=0.5,1.0,2.0$, and $4.0$ s for the lens paper.  In the inset of
Fig.~\ref{fig:pe} we show the corresponding distribution function
$P(\delta u)$. For large $\tau$'s, $P(\delta u)$ seems to be nearly
Gaussian, as expected.  For small $\tau$'s, it has an asymmetric tail
towards higher velocities, which is consistent with nonzero skewness
of the TKPZ height distribution \cite{Nee97}. We note that ash
formation increases local heat transfer ahead of the combustion front,
contributing to this effect in the velocity distribution.

\begin{figure}
  \epsfxsize=\columnwidth\epsfbox{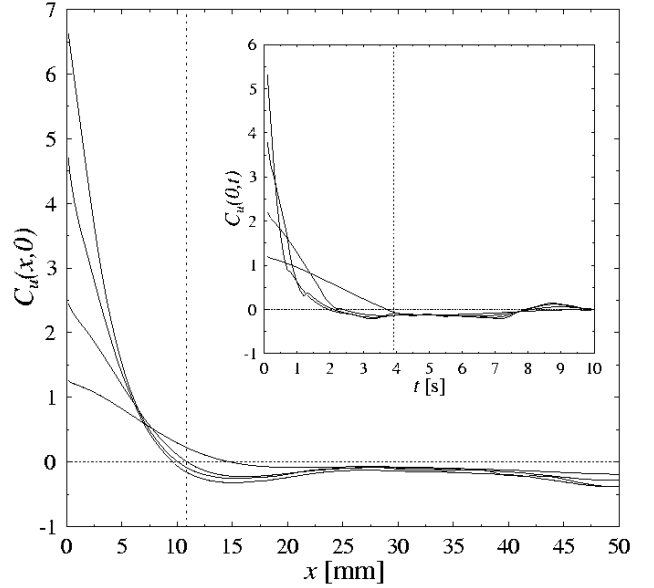}
  \caption{{The spatial and temporal velocity fluctuation correlations}
    $C_{u}(x,0)$ and $C_{u}(0,t)$ (inset), respectively, for time
    intervals $\tau=0.5,1.0,2.0,$ and $4.0$ s for the
    9.1 gm$^{-2}$ lens paper. The crossover length $r_{\rm c}$ and
    time $t_{\rm c}$ (inset) are shown by vertical lines.}
  \label{fig:vv}
\end{figure}

The two-point correlation function of these fluctuations can be
expressed in the form
\begin{equation}
  C_{u}(x,t)=\langle \delta u(x_0+x,t_0+t)\delta u(x_0,t_0) \rangle .
\end{equation}
This expression is useful for determining possible spatial and
temporal correlations. In Fig.~\ref{fig:vv} we show the spatial
correlation functions $C_{u}(x,0)$ for different time steps $\tau$,
and the corresponding temporal correlation functions $C_{u}(0,t)$ in
the inset. The effective spatial and temporal noise was found to be
uncorrelated above the crossover scales $r_c$ and $t_c$, which is in
agreement with the TKPZ behavior. Also in this case the behavior of
lens paper is analogous to that of the copier papers for which data
have been reported in Ref. \cite{Myl00}. Notice that the crossover
time is longer than the ratio of the spatial crossover scale and the
front velocity would predict.

\subsection{Universal amplitude ratios}

We now look at such properties of the TKPZ class as the universal
amplitude ratios and the universal coupling constant, discussed in the
Introduction. The experimental determination of these quantities is
slightly hampered again by the presence of the crossover scales to the
TKPZ behavior. Thus the correlation functions and the nonlinearity
$\lambda$ need to be considered in the asymptotic regime only.  This
means that the available scaling range is limited, but, as we shall
see, the results are still in rather good agreement with the
theoretical predictions. In any case, correlation functions provide
more accurate values for these quantities than the front widths, and
we therefore use them here.

\begin{figure}
  \epsfxsize=\columnwidth\epsfbox{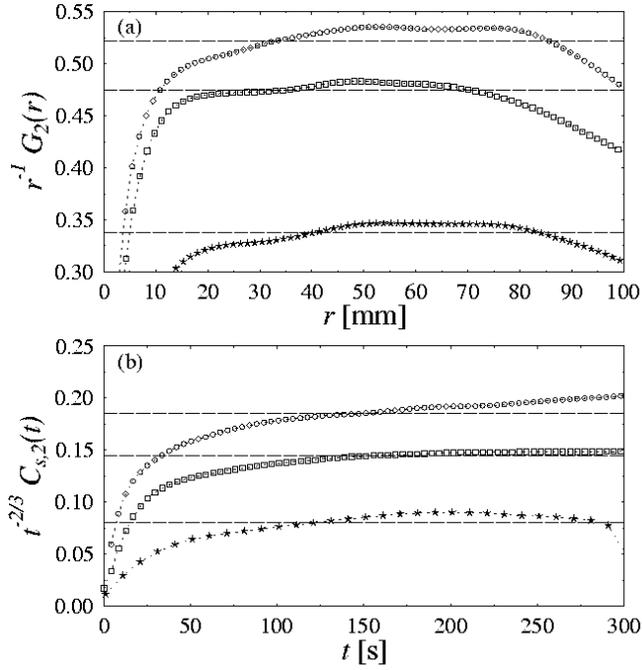}
  \caption{(a) Measured amplitude of the spatial correlation function
    $A \equiv r^{-2\chi}\,G_{2}(r)$ {\it vs.} $r$ averaged over 10
    burns of the 70 gm$^{-2}$ ($\circ$) and 5 burns of the 80
    gm$^{-2}$ copier paper ($\Box$), and 32 burns of the 9.1 gm$^{-2}$
    lens paper ($\star$) (inset). The horizontal lines indicate the
    value estimated for $A$ for different paper grades, and it has
    values $0.54$, $0.42$, and $3.47$, respectively. The roughness
    exponent was fixed at $\chi=1/2$. (b) Measured amplitude of the
    temporal correlation function $B \equiv t^{-2\beta}\,C_{{\rm
        s},2}(t)$ {\it vs.} $t$ averaged over 11 burns of the 70
    gm$^{-2}$ ($\circ$) and 18 burns of the 80 gm$^{-2}$ copier paper
    ($\Box$), and 24 burns of the 9.1 gm$^{-2}$ lens paper ($\star$)
    (inset). The horizontal lines indicate the value estimated for $B$
    for different paper grades, and it has values $0.18$, $0.15$, and
    $9.01$, respectively. The growth exponent was fixed at
    $\beta=1/3$.}
  \label{fig:AB} 
\end{figure}

We measure from the spatial and temporal correlation functions the
prefactors $A$ and $B$ (see the Introduction for a discussion).
Figure \ref{fig:AB} demonstrates the procedure by which $A$ and $B$
were determined in the saturated TKPZ regime for all three paper
grades using reasonable choices for the saturated parts of the
correlation functions.  The actual values for $A$, $B$, and the other
parameters are listed in Table \ref{tab:AB_values}. The amplitude
values were augmented by an independent measurement of $\lambda$. The
interface velocity was determined at each location with a time
interval $\tau$ long enough to attain the TKPZ regime, with a
simultaneous determination of the local interface slope. The velocity
vs. slope data were then fitted by a parabola using slopes between
-0.5 and 0.5 \cite{Alb98}, corresponding typically to about a half of
all the slopes.

Combining the results for $\lambda$, $A$, and $B$ with the exact TKPZ
exponents, we obtain for the coupling constant $g^*$ the values shown
in Table \ref{tab:AB_values}.  It is evident from this table that the
results are in good agreement with theory, with the largest deviation
showing up for the 70 gm$^{-2}$ grade. Recall that the mode-coupling
value is $g^* \simeq 0.87$ \cite{Hwa91}. If we use the measured
exponents instead, $g^*$ becomes 0.8(2), 0.8(2), and 2.7(4) for the 70
gm$^{-2}$, 80 gm$^{-2}$ and 9.1 gm$^{-2}$ grades, respectively. The
last value is due to the large effective $\beta$ for the lens paper,
caused by the very short saturated regime, and thereby poor statistics
in the temporal correlation function.  The same quantities, $\lambda$,
$A$, $B$, and the TKPZ exponents, also define the universal amplitude
ratio $R_G$, whose measured values are also shown in Table
\ref{tab:AB_values}.  As the mode-coupling and simulation results for
$R_G$ vary in the range $R_G = 0.63-0.71$, agreement with theoretical
predictions is good also in this case. Notice that $g^*$ and $R_G$ are
functionally related when the exact TKPZ exponents are used in their
expressions.

\begin{figure}
  \epsfxsize=\columnwidth\epsfbox{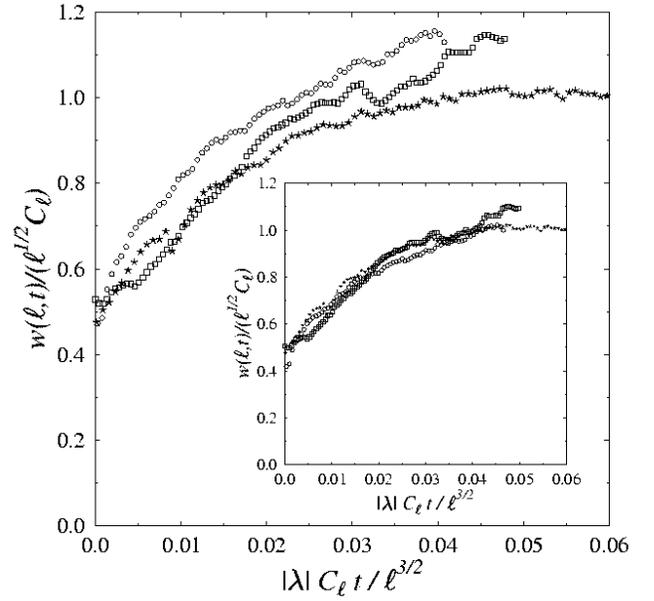}
  \caption{Scaling plots for
    $w(\ell,t)=\ell^{1/2}C_{\ell}\,F\left (\vert\lambda\vert C_{\ell}
      t/\ell^{3/2}\right )$ averaged over 39 burns of the 70 gm$^{-2}$
    ($\circ$) and 34 burns of the 80 gm$^{-2}$ copier paper ($\Box$),
    and 24 burns of the 9.1 gm$^{-2}$ lens paper ($\star$). The inset
    shows the best collapse of these data, achieved by letting
    $C_{\ell}$ vary (a variation within a few percent was enough) as a
    free parameter.}
  \label{fig:wlt} 
\end{figure}

The local width $w(\ell,t)$ allows furthermore for the determination
of a scaling function related to that of Eq.~(\ref{scal}).  One can
show, see e.g. \cite{Ama92a,Ama92b}, that $w(\ell,t) =
C_{\ell}\ell^{\chi}F(\vert\lambda\vert C_{\ell}t/\ell^z)$, where
$C_{\ell}$ is determined by the asymptotic value (at $t = \infty$) of
$w$, $w(\ell) = C_{\ell}\ell^{\chi}$.  Using a maximum value of $\ell
\simeq 14$ cm, and fitting the $C_{\ell}$ values for the three grades,
one can collapse the time-dependent data with the aid of the $\lambda$
values as measured from the slope-dependent local velocities of the
fronts \cite{Ama92a,Ama92b}.  The result is shown in
Fig.~\ref{fig:wlt}.  As the measured $\lambda$ and $C_{\ell}$ values
are not very accurate, we show in the inset of this figure the best
collapse of the data achieved by using $C_{\ell}$ essentially as a
free parameter.

%%%%%%%%%%%%%%%%%%%%%%%%%%%%%%%%%%%%%%%%%%%%%%%%%%%%%%%%%%%%%%%%%%%%%%%%%
%%                                                                     %%
%%  DISCUSSION                                                         %%
%%                                                                     %%
%%%%%%%%%%%%%%%%%%%%%%%%%%%%%%%%%%%%%%%%%%%%%%%%%%%%%%%%%%%%%%%%%%%%%%%%%
\section{Discussion and conclusions}

In this paper we have made a thorough survey and analysis of our
experiments on kinetic roughening of slow-combustion fronts in paper.
The general conclusion is that well-controlled experiments lead to a
clear asymptotic scaling that is unequivocally in accord with the
thermal KPZ universality class. This is evident from the asymptotic
behavior of the $q$th order correlation functions, and from the
scaling exponents determined from them, as well as from independent
measurements using the local width. We also augment the evidence by
computing a universal amplitude ratio characterizing the TKPZ
universality class, and the universal (KPZ fixed-point) coupling
constant, and find good agreement with theoretical estimates.

In the analysis of our experimental results it proved to be of crucial
importance to do extensive averaging over independent slow-combustion
fronts. Fluctuations in the noise affecting the fronts give rise to
wide fluctuations in individual fronts, also from burn to burn.  There
is quenched noise due to density variations typical of paper-like
materials, and the effective noise also includes dynamical effects. A
burning front creates around it a fluctuating flow of air, and this
flow will give rise to fluctuations in the ``effective'' heat
conductivity (due to convective transfer of heat) and in cooling.
These dynamical fluctuations could be demonstrated by adjusting the
overall air flow in the combustion chamber, and thereby regulating the
crossover scales parallel to the flow. The temporal crossover scales
are always clearly longer than the spatial crossover scales along the
fronts divided by the average front velocities. This is indicative of
dynamical convective effects. We found that averaging over
approximately ten independent burns is needed for fairly reliable
estimates for the quantities measured. For copier paper this means in
practice averaging over about 40000 individual fronts, and for lens
paper, with much faster propagating fronts, over several thousand
individual fronts. For the lens paper, averaging was hampered by the
shortness of the saturated regime.

In addition to the demonstration of the asymptotic TKPZ behavior,
another interesting feature is the persistent crossover present in
slow combustion of paper. The apparent scaling properties of the
fronts are markedly different at short time and length scales as
compared to the asymptotics. This phenomenon was first observed
\cite{Hor91a} in fluid-flow experiments, and very recently also in
\cite{Sur99} penetrating flux fronts in thin-film superconductors. As
discussed in the Introduction, several possible mechanisms have been
suggested for this phenomenon (see also Ref.~\cite{Hal95} for a
review).  Before our work, the reasons and consequences of the two
regimes was not settled experimentally though power-law distributed
amplitudes in the effective noise were reported both in \cite{Hor91b}
for fluid-flow experiments and in \cite{Myl00} for slow combustion of
paper.  In the latter experiments, performed by us, noise correlations
were furthermore found to be short ranged, both in space and in time.
Moreover, the ``decay lengths'' of these correlations seem roughly to
coincide with the crossover scales in the height-height correlation
functions. It thus appears that the SR correlations in the noise
affecting the fronts, either quenched or annealed, or both, are likely
to be responsible for the crossovers and the related higher apparent
exponents at short range. The continuous decay of SR correlations
would also explain the lack of true scaling in that regime,
\textit{i.e.} the running exponents do not show plateaux there. In the
case of slow combustion of paper at least, the effective noise clearly
is partly of dynamical origin.  Thus quenched noise alone cannot be
decisive for the short-range effects, although it leads to a similar
crossover behavior. Our measured SR exponents shown in
Table~\ref{tab:res_2} do not agree with the apparent SR exponents for
moving fronts with quenched noise: close to the depinning transition
$\beta\simeq\chi\simeq 0.75$ \cite{Les96}. Some effects at SR due to
the irrelevant terms in the appropriate Langevin equation cannot of
course be ruled out.

Multiscaling at short range seems, on the other hand, to result from
other features in the fronts. It is evident that regions of high
gradient (in absolute value) are amplified in the higher-order
correlation functions, and thereby affect the multiscaling properties
of the fronts. Distinct regions of such high gradient values appear if
successive fronts are plotted as surface diagrams. Note that
digitizing errors, and sharp natural defects in the observed data,
also make a contribution to the apparent multiscaling properties, and
should be filtered out as artefacts. Fairly sharp steps, which
occasionally appear in the (filtered) fronts, begin to move also
sideways, driven by the nonlinear term in the KPZ equation
Eq.~(\ref{KPZ}). These moving structures appear as local
``avalanches'' in the propagating fronts. By analyzing distinct
intervals of propagating combustions fronts, containing varying
amounts of such avalanches, we can conclude that the $q$-dependence at
SR of the slopes of the spatial and temporal correlation functions
extend longer when there are more avalanches present. This effect is
however difficult to quantify precisely as it is not possible to
completely remove the avalanches, nor the experimental artefacts, from
real data.

We have only considered here well-propagating fronts in which pinning
effects do not appear. By reducing the KNO$_3$ concentration, we can
however approach the pinning regime in a rather controlled way. A lot
of statistics must be produced again, since the exact amount of
KNO$_3$ absorbed in the paper samples (the ``driving force'') is
difficult to regulate. Nevertheless, in this way we think one can
eventually determine the interface behavior also at pinning.  It is
evident that there will also be crossovers present in this case, with
the actual crossover scales probably depending on the driving force.

Acknowledgements: This work has been in part supported by the Academy
of Finland under the MATRA Program and the Center of Excellence
Program (Project No. 16498).

%%%%%%%%%%%%%%%%%%%%%%%%%%%%%%%%%%%%%%%%%%%%%%%%%%%%%%%%%%%%%%%%%%%%%%%%%
%%                                                                     %%
%%  REFERENCES                                                         %%
%%                                                                     %%
%%%%%%%%%%%%%%%%%%%%%%%%%%%%%%%%%%%%%%%%%%%%%%%%%%%%%%%%%%%%%%%%%%%%%%%%%
%
% now the references. delete or change fake bibitem. delete next three
%   lines and directly read in your .bbl file if you use bibtex.
%
%-------------------------------------------------------------------------

%%%%%%%%%%%%%%%%%%%%%%%%%%%%%%%%%%%%%%%%%%%%%%%%%%%%%%%%%%%%%%%%%%%%%%%%%
%%                                                                     %%
%%  FIGURES                                                            %%
% \begin{figure}
% \caption{}
% \label{}
% \end{figure}
%
%-------------------------------------------------------------------------

\pagebreak

\pagebreak
% 
%%%%%%%%%%%%%%%%%%%%%%%%%%%%%%%%%%%%%%%%%%%%%%%%%%%%%%%%%%%%%%%%%%%%%%%%%
%%                                                                     %%
%%  TABLES                                                             %%
% \begin{table}
% \caption{}
% \label{}
% \end{tabular}
% \end{table}
%
%------------------------------------------------------------------------
%
%\begin{table}[h]
%\caption{Paper grades used in experiments with basis weights and 
%average propagation velocities.}
%\begin{tabular}{l|c|c}
%Paper grade & Basis weight & Average velocity \\ \hline
%Copier paper 1 & 70~g\,m$^{-2}$  & 0.64(2)~mm\,s$^{-1}$ \\
%Copier paper 2 & 80~g\,m$^{-2}$  & 0.50(2)~mm\,s$^{-1}$ \\
%Lens paper     & 9.1~g\,m$^{-2}$ & 9.0(5)~mm\,s$^{-1}$  
%\end{tabular}
%\end{table}
\pagebreak
\begin{widetext}
\begin{table}[b]
        \caption{The scaling exponents $\beta$ and $\chi$, 
        crossover values $r_c$ and $t_c$, and average
        velocities.}
        \protect{\label{tab:res}}
\begin{tabular}{l|ddd|ddd}
                & \multicolumn{3}{c|}{$C_{2}(r,t)$}
                & \multicolumn{3}{c}{$w(\ell,t)$}    \\
                & 70~g\,m$^{-2}$  & 80~g\,m$^{-2}$  & 9.1~g\,m$^{-2}$  &%
                 70~g\,m$^{-2}$  & 80~g\,m$^{-2}$ & 9.1~g\,m$^{-2}$   %
\\ \hline
$\chi_{SR}$     & 0.88(2)         & 0.89(2)         & 0.83(1)        &%
                  0.81(6)         & 0.83(5)         & 0.81(1)        \\
$\chi_{LR}$     & 0.53(3)         & 0.51(3)         & 0.53(4)        &%
                  0.57(1)         & 0.55(2)         & 0.56(2)        \\ \hline
$r_c$ [mm]      & 4.7(4)          & 6.0(5)          & 11(2)          &%
                  12(7)           & 14(6)           & 18.8(4)        \\ \hline
$\beta_{SR}$    & 0.59(4)         & 0.69(2)         & 0.61(2)          &%
                  --              & --              & --             \\ 
$\beta_{LR}$    & 0.40(3)         & 0.39(3)         & 0.46(2)          &%
                  0.29(3)         & 0.32(3)         & 0.28(5)        \\ \hline
$t_c$ [s]       & 25(9)           & 27(5)           & 3.7(4)           &%
                  --              & --              & --             \\ \hline
$v$ [mm/s]      & 0.64(2)         & 0.50(2)         & 9.0(5)           &%
                                  &                 & 
\end{tabular}
\end{table}
\begin{table}[b]
        \caption{The scaling exponents $\beta$ and $\chi$ 
        obtained by first subtracting the intrinsic widths from the data.}
        \protect{\label{tab:res_2}}
\begin{tabular}{l|ddd|ddd}
                & \multicolumn{3}{c|}{$C_{2}(r,t)$}
                & \multicolumn{3}{c}{$w(\ell,t)$}    \\
                & 70~g\,m$^{-2}$  & 80~g\,m$^{-2}$  & 9.1~g\,m$^{-2}$  &%
                 70~g\,m$^{-2}$  & 80~g\,m$^{-2}$ & 9.1~g\,m$^{-2}$   %
\\ \hline
$\chi_{SR}$     & 0.90(3)         & 0.90(4)         & 0.85(1)          &%
                  0.84(6)         & 0.87(8)         & 0.95(6)          \\
$\chi_{LR}$     & 0.50(4)         & 0.47(4)         & 0.50(6)          &%
                  0.56(1)         & 0.52(5)         & 0.51(1)      \\ \hline
$\beta_{SR}$    & --              & 0.75(5)         & 0.64(3)          &%
                  --              & --              & --           \\ 
$\beta_{LR}$    & 0.36(3)         & 0.34(4)         & 0.43(6)          &%
                  --              & --              & --           
\end{tabular}
\end{table}
\begin{table}[h]
        \centering
        \caption{Results for the correlation amplitudes, $\lambda$, and the 
                 universal quantities using $\beta=1/3$ and $\chi=1/2$.}
        \label{tab:AB_values}
        \begin{tabular}{l|d|d|d|d|d}
  Paper grade    &  $A$   &  $B$      &  $\lambda$  & $R_{G}$ & $g^{*}$ %
  \\ \hline
    70 gm$^{-2}$  & 0.52(2) & 0.186(12) & 0.465(2)  & 0.74(6) & 0.79(9)    \\ 
    80 gm$^{-2}$  & 0.475(7)& 0.14(1)   & 0.370(1)  & 0.73(5) & 0.76(8)    \\ 
    9.1 gm$^{-2}$ & 3.4(1)  & 8.0(8)    & 4.0(4)    & 0.62(8) & 1.0(2)        
        \end{tabular}
\end{table}
\end{widetext}
\end{document}